\documentclass[10pt,conference]{IEEEtran}

\usepackage{color}

\usepackage[pdftex]{graphicx}

\IEEEoverridecommandlockouts

\begin{document}

\title{FFT-based Network Coding For Peer-To-Peer Content Delivery\thanks{This work was supported by the French ANR grant No 2006 TCOM 019 (CAPRI-FEC project).}}

\author{
\IEEEauthorblockN{Alexandre Soro and J\'er\^ome Lacan}
\IEEEauthorblockA{University of Toulouse,\\ 
ISAE/DMIA,\\
Toulouse, France\\ 
\{alexandre.soro,jerome.lacan\}@isae.fr}
}

\maketitle

\begin{abstract}

In this paper, we propose a structured peer-to-peer (P2P) distribution scheme based on Fast Fourier Transform (FFT) graphs. We build a peer-to-peer network that reproduces the FFT graph initially designed for hardware FFT codecs. This topology allows content delivery with a maximum diversity level for a minimum global complexity. The resulting FFT-based network is a structured architecture with an adapted network coding that brings flexibility upon content distribution and robustness upon the dynamic nature of the network. This structure can achieve optimal capacity in terms of content recovery while solving the problem of last remaining blocks, even for large networks.

\end{abstract}

\section{Introduction and Related Work}

The Internet has seen the emergence of new ways of communications and services these recent years. Peer-to-peer streaming \cite{pplive} \cite{sopcast} \cite{ppstream} \cite{tvants} and content delivery \cite{bittorrent} solutions have shattered the classical client-server model. This approach brings new technical challenges, in particular efficient methods to deliver content to all clients regarding the bandwidth used and time constraints.

The problem of a fair use of the bandwidth of each client has always been a critical point in a P2P based network. The first refinements over a classical client-server approach used replication schemes. By duplicating blocks between a subset of the fastest clients \textit{i.e.} nodes, the average bandwidth used is improved. However this mechanism suffers from two main disadvantages, as the fastest nodes become critical in the network, and also that the uplink of the other nodes is unused.

Structured and hierarchical topologies can also be designed for content delivery. The network can be built with a single tree-based approach \cite{Banerjee02scalableapplication} or with a more sophisticated multi tree-based approach \cite{Nicolosi03p2pcast:a} \cite{Castro03splitstream:high-bandwidth}. In this case, in any branch of the tree, a node, usually the fastest one, is designated to be the head of the subset and represents the branch to interact with the upper layers. This node is also responsible for delivering the content from the upper layers to the other nodes of the subset. However, this scheme suffers from the same problem than the replication approach in a P2P scope as this node can leave at any time, leaving the other nodes and the lower layers orphans. In order to have more reliable connections, multi-head schemes have also been investigated \cite{DBLP:journals/jsac/TranHD04}, with delivery and representation to upper layers separated. Nevertheless, this solution remains more complex to build.

In order to improve these points, a new class of unstructured techniques, has appeared, based on the applications of coding theory \cite{MWSl77} over P2P networks. Depending on the application target, multimedia chunks or data content can be split into blocks. Compared to a classical P2P application, the coding theory offers the possibility to introduce new blocks, which are built from the existing blocks, as linear combinations, in the network. These new blocks help provide diversity which is important in a P2P network, as a client only gets a local view of the network. This diversity also brings improvements on load balancing as the scope of clients, of which a node may be interested in, is increased. Two main families appear using coding theory. In the first one, which is commonly known as \textit{erasure coding}, a server is responsible for creating new redundant blocks, which are then disseminated in the network like source blocks \cite{dairaine:ComputerNetwork}. 
The more redundant blocks are created, the more diversity is spread in the network. A good solution to improve the diversity in the network is to use \textit{network coding} which allows intermediate nodes to perform coding operations. With this approach, the diversity is created by each client which mimics the behaviour of the server with blocks already received \cite{Gkantsidis05Network}. It has to be noticed that as the linear combinations of existing blocks are created with no special rule, the coefficients and indices of the linear combinations have to be embedded in the block header. However, because of the lack of structure, network coding can not guarantee that the decoding is always possible with a minimum number of received blocks which implies extra blocks to be transmitted.

In our paper, we propose a new structured peer-to-peer architecture using network coding, which is based on the application of a Fast Fourier Transform (FFT) computation scheme. This scheme has the advantage to be of optimal diversity while requiring a minimum set of blocks. This topology is also more flexible than classical structured topologies, while being more reliable in a P2P scenario where any node can leave at any time. Rather than focusing on some practical issues, we discuss on how some properties of the FFT graph can bring significant improvements over existing P2P networks.

In Section \ref{FFT} we present the FFT in a particular finite field and some of its properties regarding a P2P network. Examples of application cases will be discussed in Section \ref{apps}. In Section \ref{conclusions}, we draw some conclusions and perspectives for the FFT-based network coding.

\section{FFT-based Network Coding}
\label{FFT}

\subsection{Presentation of the FFT network}

In this paper, as we deal with network coding, a data file $x$ is divided in $k$ blocks $x_0,x_1,...,x_{k-1}$, which are the basis to create new blocks by linear combinations. Each client of the P2P network receives linear combinations of these $k$ blocks. Hence, the client can recover the initial data file $x$ if and only if the associated matrix of the linear combinations received is of rank $k$. Optimal codes from this point of view are Maximum Distance Separable codes (MDS) \cite{MWSl77} as any subset of $k$ coefficients (\textit{i. e} of $k$ linear combinations) of a codeword  allows to recover the $k$ source coefficients (\textit{i. e} is of rank $k$). Among these codes, a well-known class of codes are Reed-Solomon codes whose the encoding process can be viewed as the Discrete Fourier Transform (DFT). Let the Galois field $GF(q)$ be the finite field with $q$ elements and let $\omega$ be an element of $GF(q)$ of order $n$ (\textit{i. e} $\omega^{n} \equiv 1$ and $\omega^j \neq 1$ for $0<j<n$). The Discrete Fourier Transform is then defined as follow :

Let $a$ be a vector of size $n$, $a=(a_0,a_1,...,a_{n-1})$. Then the DFT of $a$ is the vector $A$ of size $n$, $A=(A_0,A_1,...,A_{n-1})$ with :

$$
A_{i} = \sum_{j=0}^{n-1}{a_{j} \omega^{ij}}
$$

It is also possible to define the Inverse Discrete Fourier Transform (IDFT), which is then defined by :

$$
a_{i} = \frac{1}{n}\sum_{j=0}^{n-1}{A_{j} \omega^{ij}}
$$

The Fast Fourier Transform (FFT), and its derivates, is the fastest method to compute the DFT - and the IDFT, and was introduced by Cooley and Tukey \cite{Cooley65}. When $n$ is a power of 2, the complexity of computing the DFT falls to $O(n\log n)$. This improvement over the classical quadratic complexity comes from a \textit{divide-and-conquer} approach. Indeed, FFT reduces the processing of the DFT of size $n$ to a problem of processing two DFTs of size $n/2$ and a few operations recursively. The objective of this paper is to build an FFT network that ensures optimal diversity at the lowest complexity cost.

The first step is to find a finite field containing elements of order $2^u$. Indeed, the finite field $GF(2^m)$ classically used to build erasure or error correcting codes does not contain such element. For this reason, like in \cite{blahut:1985}, we choose to process the FFT on the field $GF(65537)$. Indeed, the primitive roots of this field are of order $65536=2^{16}$ and it is easy to build some elements of order $2^u$, for $1\leq u \leq 16$. Working on this field allows fast computations since they are processed on integers modulus $65537$. Moreover, these coefficients can be easily stored in data blocks or packets  as described in \cite{ccnc}. Remarking that $65537$ is a Fermat prime, in this case, the FFT in this field can also be called a Fermat Number Transform (FNT). In this field, the adaptation of the FFT is straightforward and allows size up to $2^{16}$ blocks.

\begin{figure}[htb]
  \begin{center}
   \includegraphics[width=.4\textwidth]{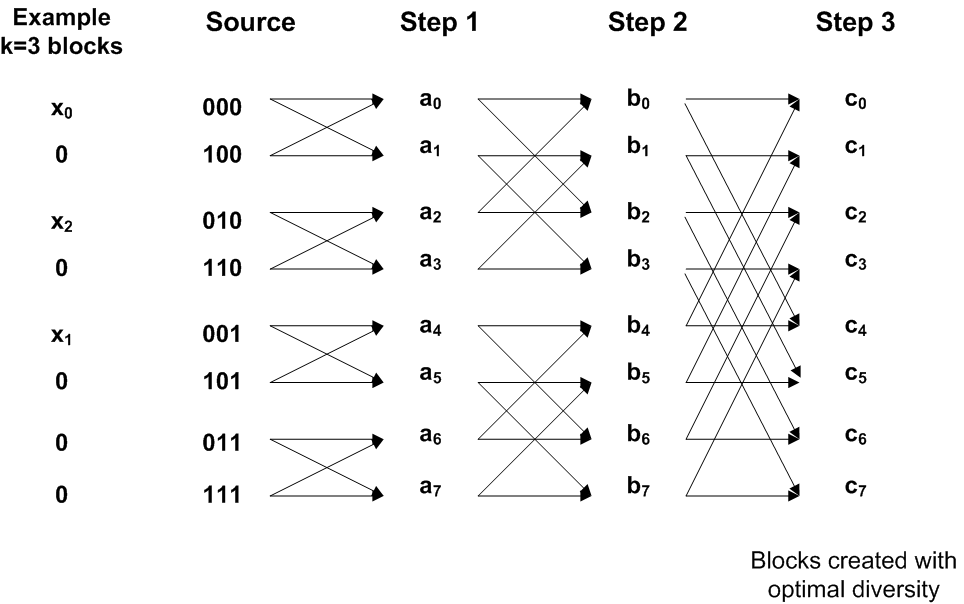}
    \end{center}
    \caption{The FFT computation graph for 8 elements}
   \label{fig:fft}
\end{figure}

In the general way, FFT computation and its recursive form can be represented as a graph as seen in Fig. \ref{fig:fft}. As expected, the FFT computation graph has $\log n$ steps. In order to match the process of the FFT, some precisions have to be done on the source positions and the computations of the non-source positions. In fact, thanks to the FFT algorithm \cite{Cooley65}, when an FFT of size $n$ is divided in the processing of 2 FFTs of size $n/2$, the source positions are separated by even and odd values. Recursively, it has the consequence to change the way the source positions are placed in the FFT computation graph. Indeed, in a FFT graph, the source of indice $i$ will have the position $j$ where $j$ is the reversed binary representation of $i$. In addition, the computations resulting from the recursivity impact on the computations of each non-source node. From this point, we represent each node of the FFT graph by two coordinates, its step, which corresponds to its distance to the source positions and its position in its step $(s,p)$. As a node receives information from two sources of nodes of the previous step $(i_1,i_2)$, the FFT gives the following computation for a node $N(s,p)$:

$$
N(s,p) = N(s-1,\alpha_p) + \omega^{(np 2^{-s}) mod\ n}\times N(s-1,\alpha_p+2^{s-1})
$$

with 

$$
\alpha_p = 2^s\times \lfloor \frac{p}{2^s} \rfloor + p\ mod\ 2^{s-1}
$$

With the constraints defined above, it is possible to directly map the nodes of a peer-to-peer network to the nodes of the FFT computation graph. The structured connections mapping between all the clients will follow the connections between the intermediate nodes of the graph. From this point, we will consider nodes of the FFT computation graphs as clients of a P2P topology. In this representation, the blocks of the servers are in fact the source positions of the FFT \textit{i.e} the positions of step 0.

In general, the number of blocks $k$ is not a power of 2, meaning that the FFT graph of size $n$ will be the smallest power of 2 greater than $k$. In this case, the last $n-k$ indices of the source are null and the corresponding positions and links are void in the graph, according to the reversed binary positioning.

The FFT structure brings many advantages in a P2P network. At the last step, $n$ blocks are processed from the source vector $x$ of size $k$. As we follow the FFT algorithm, in fact, these $n$ blocks are the FFT of the vector $x$. It is possible to represent these linear combinations by a $k\times n$ matrix. Remarking that this matrix, which is the matrix of the FFT, is a special case of a Vandermonde matrix with consecutive powers of $\omega$, it is then the generator matrix of a MDS Reed-Solomon code which provides optimal diversity, as it is possible to recover the source data $x$ with any subset of $k$ elements of the last step, as previously introduced. In other words, any $k$ linear combinations in the last step are of rank $k$. It is also important to notice that this diversity is created in a logarithmic number of steps. In addition, using an optimized technique \cite{ccnc} on nodes of the last step allows the source data to be also decodable with the same order.

To illustrate this property, in Fig. \ref{fig:fft}, we have represented a case with $k=3$ source blocks and a FFT of size $n=8$. At the last step, $8$ blocks have been created according to the FFT. If we take the nominal case where a client connects only to nodes of the last step, the FFT coding is optimal, in the sense that a client can decode the initial $3$ source blocks $(x_0,x_1,x_2)$ with any $3$ blocks of the last step.

This network can be applied to a wide area of applications from distributed content or secret sharing to P2P streaming applications. However, the principles and the requirements of these applications are also different. From this point, we will introduce several mechanisms of the FFT-based network coding that may apply to some of these scenarios.

\subsection{Services offered by the FFT network to the clients}

As explained previously, the main result of the P2P network are the blocks produced at the last step nodes of the network. However, in order to improve the global service provided by the FFT network, it could be interesting for a client, which may or may not be part of the FFT graph, to have the possibility to decode from the blocks produced by the intermediate nodes also. These intermediate nodes produces linear combinations of a part of the source blocks only, so the possibility to decode with only $k$ blocks is not guaranteed anymore. However, thanks to the structure of the network, and the linear combinations induced, it is instantly possible for a client to determine if a node can bring useful information to it or not. This is an important point as any encoded block that will transit in the network is an innovative block for its receiver, and no bandwidth is lost in uninteresting blocks. 

\subsection{Flexibility}

It is also possible for a node to take virtually place on several positions in an FFT network. This point may be justified by higher capacity nodes in terms of throughput, and more generally by load balancing in the network. This may be also an answer to the duplication scheme described in the introduction, as sharing the content between several nodes is relevant. However, in this solution, we avoid the problem of critical nodes and we also allow the uplink of each client to be used. To this point, each node in the network has been treated equally in terms of capacity. However, for example, in an Internet based scenario, throughputs are variable from one connection to another. And, because of the structure of the FFT graph, the network does not really benefit from faster links, as they have to wait from slower ones. By allowing some terminals to play the role of several nodes in the graph, we can take advantage of these differences by approaching the capacity of these faster nodes. Moreover, it also reduces the number of real connections between nodes and then the data transferred in the network.

Given a terminal which replaces $m$ nodes in an FFT network, the point is to determine how to choose these $m$ points in order to minimize the overall traffic. Hence, this problem is equivalent to find a combination of $m$ elements in the FFT graph, such that the number of connections between these $m$ nodes is maximal. For $m=4$, the case where the four nodes are taken from a closed cross can be viewed as optimal. If $m_i$ represents the $i^{th}$ node, a closed cross brings the following four connections : $m_1-m_3$,$m_1-m_4$,$m_2-m_3$ and $m_2-m_4$. It is more than the three connections induced by arbitrary consecutive nodes. This scheme is also the representation of the FFT of size 2. More generally, if $m$ is of the form $m=j\times2^{j-1}$, the optimal repartition of these points is to form an FFT of size $2^{j-1}$. If $m$ does not fit in an FFT, an optimal construction is a subset of the smallest FFT of a bigger size $s=2^j$. In this case the first $2^{j-1}$ points of $m$ will form an FFT and the remaining points will form FFTs of smaller sizes, where all these FFTs will be connected.

\subsection{Reliability}

If the approach presented above is sufficient when the nodes of the FFT graph are static, it may not be sufficient for a P2P scheme where any node can leave at any time. The behaviour of the network has to be defined when a node wants to join or leave the network. If a node joins the network, the question is to put this node in the FFT network, and also, to enlarge the network, if there is no space available. If a node leaves the network, its position has to be replaced, but also, the network has to keep a minimal size. In addition, the number of changing connections has to be minimal in order to keep the signalling flow low compared to the data traffic. Finally, any change in the network structure has to be synchronized in the whole network. 

An FFT of size $n$ is full when all its non-source nodes that can be connected are used and all different in a dynamic configuration. In the case where $k=n$ , the number of non-source nodes in the network is, for example, $2^{n-1}(n-1)$. When a node wants to join a FFT-based network, if the network is not full, then the node takes the first available place in the graph following the next procedure. If the node joins a full FFT network, then a new stage has to be created, respecting the positions and the configuration of the FFT.

\begin{figure}[htb]
  \begin{center}
   \includegraphics[width=.4\textwidth]{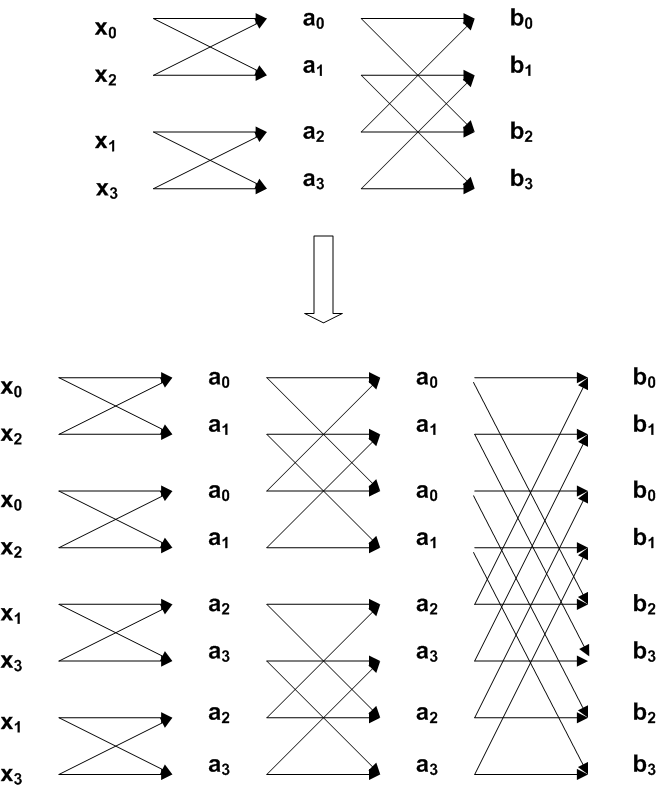}
    \end{center}
    \caption{Example of FFT extension from 4 blocks to 8}
   \label{fig:rep}
\end{figure}

First, for this, we introduce what we call an \textit{extension technique}. This technique is depicted in Fig. \ref{fig:rep} where an initial graph with two steps is transformed into a graph with three steps. In fact, this extension is purely a graph view. The number of blocks becomes twice, but the throughput of each becomes half. We can see that the connections between step 1 and step 2 in the new graph are between the same nodes, so no traffic transits in this state. As each node is replicated twice, the global throughput for each node is still the same. In addition, thanks to this approach, no connections are changed, but only the operations associated with these nodes. Then, after the extension process, the node that wants to join takes the position of the first replicated node.

In the same manner, in a dynamic configuration, any node can also leave the network at any moment. It means that if a node that is not located at the last stage of the FFT leaves, its position has to be replaced. This node is always connected to two nodes of the upper step and also two nodes of the lower step at least. Possible replacement policies are discussed in the following. These connections allow a greater reliability, for example, than a tree-based approach, when a node of the main branch leaves. This mechanism is then quite close to an unstructured approach where a node is, in practice, not connected to an infinity of nodes either.

As the behaviour of each node is unpredictable and independant, there is no reason to give priority to one policy. Hence, the most straightforward way to replace the leaving node, is to choose the last non-replicated available node. It is possible to then replace this position in order to prepare for an eventual reduction of the FFT network. The reduction of the FFT network is possible when all connected nodes can fit in an FFT network of smaller size. This operation is the inverse of the replication process and detailed in Fig. \ref{fig:rep}. This operation also keeps the connections and the global throughputs between the nodes unchanged.

Another benefit of the FFT network is that upon disconnection of a source, the flow is not interrupted and decoding is still possible seamlessly. On the server side, the number of blocks can also be adapted by adding or removing blocks. In the case of a full FFT where the decoding is possible with $k$ of $n$ end terminals, for example, during this transition, the decoding is still possible with $k$ nodes and one of the decoded information will simply be zero. At the end of the transition, the decoding becomes possible with only $k-1$ nodes.

\section{Application to peer-to-peer and classical scenarios}
\label{apps}

\subsection{Peer-to-peer based applications upon dynamic configuration}

Peer-to-peer applications and especially streaming contexts arise several requirements and issues. We try to enumerate some crucial points and how an FFT-based topology could answer these topics.

\begin{itemize}
 \item \textit{Dynamic context} : In a scheme where any node wants to leave or join at any time, the network has to adapt to these behaviours at any time. As we have seen, mechanisms such as extension technique bring an answer when many peers wants to join a network, which is often the case at the beginning of a transfer or a broadcast. In the same manner, the FFT network is able to adapt itself if a node leave or at worst, if its size has to be reduced. This reduction mechanism is also very important and has to be used as soon as possible, as it also implies a reduction of the complexity of the global network. In addition, this network can support a dynamic load of the sources, if more or less blocks have to be transmitted, and this, without any changes to the clients.
 \item \textit{Heterogeneity of the peers} : Another important point is a good use of the ressources of each peer. Suited to the heterogeneity of each peer, the FFT network provides the possibility of virtual nodes, which allows faster nodes to have a greater importance in the network in terms of bandwidth. With this mechanism based on FFT of different sizes, it is also possible to achieve a fine tuning, close to the limits of each client, if the ressources of each node is known in advance.
 \item \textit{Limited bandwidths and connections} : As the bandwidth and the number of clients one node can connect to is limited, and thus for several different reasons : physical layer, geographical situation... the \textit{a priori} knowledge of the useful nodes for decoding the source unit is an innovative point. It implies that no useless connections are made and consequently, it has an impact on the connections delay, which is a decisive point in streaming networks.
 \item \textit{Selfish behaviour} : As a member of the FFT network is always connected to two nodes of greater steps, at least, it is easy to determine if one node does not actively participate and thus signalling it to the other nodes. However, counter-measures against non FFT network members remain an open point.
\end{itemize}

\subsection{Distributed backup and content delivery upon static configuration}

In this context, the hypothesis is taken that the nodes of the FFT network are static or quasi-static. The intermediate nodes can be either the same nodes as the source nodes or different. In the case of a distributed backup, when the data of $k$ terminals has to be replicated to $n$ terminals with the guarantee that all the data can be accessed by any subset of $k$ terminals, the FFT based network can be used with several improvements upon Reed-Solomon based methods. In this case, the intermediates nodes are in fact also the $n$ terminals and the data is transmitted within these terminals. In the general case, $n$ is not a power of 2. Hence, the missing intermediate nodes have to be replaced by the existing terminals. In an homogeneous network, if it is possible to choose arbitrarly which existing terminals will replace the missing nodes, it may be useful to use the FFT repartition provided above for each terminal in order to minimize the data transferred in the network.

Among the benefits of the FFT topology, first, as the structure of the network is regular, the network is more fault tolerant than an irregular structure as no node is more critical than another. In addition, this regular structure allows a better adaptability to heterogeneous terminals. Indeed, as it is possible to dedicate several virtual nodes to one real terminal, mapping each terminal to a certain number of virtual nodes, depending on their relative speed, helps to get as close as possible to an optimal delivery solution in terms of bandwidth optimization for each terminal.

This solution is also optimal in terms of decoding speed $O(n\log n)$ using an adapted decoding \cite{ccnc}. When the number of sources is greater than several hundreds, this complexity allows far greater speeds ($\simeq$ 10Mbps) than the classical quadratic solutions of existing MDS codes. It has not to be forgotten that the FFT-based codes also support quadratic decoding complexity algorithms which allows speeds of more than 100Mbps for up to 128 sources.


Then, the FFT can be considered as the first solution for distributed backup and content delivery of optimal code capacity, that can be deployed for networks upon thousand of terminals, with logarithmic encoding and decoding complexities. 

\section{Conclusions}
\label{conclusions}

In this paper, we have proposed an efficient structured peer-to-peer topology, relying on the use of a network coding scheme based on a FFT graph. This network produces at its last step a set of blocks with an optimal level of diversity for a minimal global complexity. Moreover, in this structure, each client has a global virtual view of the whole network, in the sense that every client is able to determine from which node he has to download content, in order to decode the source transmitted by servers. This structure also guarantees an excellent reliability as each node is always connected to several other clients at any time. Moreover, the diversity and the fact that all transmitted blocks have to be useful, implied by the FFT, makes it an excellent candidate for an efficient global content delivery, even for large structures.

Many particular applications can benefit from this coding, like peer-to-peer TV (P2PTV) or content sharing applications. This structure addresses the problem of the local view of a client in existing applications, where a node can wait for some time for a missing block, which is available elsewhere, for example. Even if this topology is built for dynamic nodes, this paper also brings an innovative solution for static mechanisms such as distributed content backup, where an optimal recovering capacity is required, and for a minimum software cost.

In a future work, we plan to define the practical characteristics of a FFT-network, such as the address representation of the nodes, the signalling between the nodes, and the communication protocol. The objective is first to evaluate the gain over classical peer-to-peer approaches and then to implement the FFT structure in real conditions in order to obtain a peer-to-peer network with optimal properties in terms of diversity, reliability and flexibility.

\bibliographystyle{IEEEtran}
\bibliography{fft_p2p}

\begin{thebibliography}{10}
\providecommand{\url}[1]{#1}
\csname url@rmstyle\endcsname
\providecommand{\newblock}{\relax}
\providecommand{\bibinfo}[2]{#2}
\providecommand\BIBentrySTDinterwordspacing{\spaceskip=0pt\relax}
\providecommand\BIBentryALTinterwordstretchfactor{4}
\providecommand\BIBentryALTinterwordspacing{\spaceskip=\fontdimen2\font plus
\BIBentryALTinterwordstretchfactor\fontdimen3\font minus
  \fontdimen4\font\relax}
\providecommand\BIBforeignlanguage[2]{{%
\expandafter\ifx\csname l@#1\endcsname\relax
\typeout{** WARNING: IEEEtran.bst: No hyphenation pattern has been}%
\typeout{** loaded for the language `#1'. Using the pattern for}%
\typeout{** the default language instead.}%
\else
\language=\csname l@#1\endcsname
\fi
#2}}

\bibitem{pplive}
\BIBentryALTinterwordspacing
``Pplive.'' [Online]. Available: \url{http://www.pplive.com}
\BIBentrySTDinterwordspacing

\bibitem{sopcast}
\BIBentryALTinterwordspacing
``Sopcast.'' [Online]. Available: \url{http://www.sopcast.com}
\BIBentrySTDinterwordspacing

\bibitem{ppstream}
\BIBentryALTinterwordspacing
``Ppstream.'' [Online]. Available: \url{http://www.ppstream.com}
\BIBentrySTDinterwordspacing

\bibitem{tvants}
\BIBentryALTinterwordspacing
``Tvants.'' [Online]. Available: \url{http://www.tvants.com}
\BIBentrySTDinterwordspacing

\bibitem{bittorrent}
\BIBentryALTinterwordspacing
``The bittorrent protocol specification.'' [Online]. Available:
  \url{http://www.bittorrent.org/beps/bep\_0003.html}
\BIBentrySTDinterwordspacing

\bibitem{Banerjee02scalableapplication}
S.~Banerjee, B.~Bhattacharjee, and C.~Kommareddy, ``Scalable application layer
  multicast,'' 2002.

\bibitem{Nicolosi03p2pcast:a}
A.~Nicolosi, ``P2pcast: A peer-to-peer multicast scheme for streaming data,''
  in \emph{1st IRIS Student Workshop (ISW’03)}, 2003.

\bibitem{Castro03splitstream:high-bandwidth}
M.~Castro, P.~Druschel, A.-M. Kermarrec, A.~Nandi, A.~Rowstron, and A.~Singh,
  ``Splitstream: High-bandwidth multicast in cooperative environments,'' 2003,
  pp. 298--313.

\bibitem{DBLP:journals/jsac/TranHD04}
D.~A. Tran, K.~A. Hua, and T.~T. Do, ``A peer-to-peer architecture for media
  streaming,'' \emph{IEEE Journal on Selected Areas in Communications},
  vol.~22, no.~1, pp. 121--133, 2004.

\bibitem{MWSl77}
F.~I. MacWilliams and N.~J.~A. Sloane, \emph{The Theory of Error-Correcting
  Codes}.\hskip 1em plus 0.5em minus 0.4em\relax North-Holland, 1977.

\bibitem{dairaine:ComputerNetwork}
L.~Dairaine, L.~Lanc\'erica, J.~Lacan, and J.~Fimes, ``{Content-access QoS in
  peer-to-peer networks using a fast MDS erasure code},'' \emph{Computer
  Communications}, vol. vol. 28, no.~15, pp. 1778--1790, September 2005.

\bibitem{Gkantsidis05Network}
C.~Gkantsidis and P.~R. Rodriguez, ``Network coding for large scale content
  distribution,'' in \emph{INFOCOM 2005. 24th Annual Joint Conference of the
  IEEE Computer and Communications Societies. Proceedings IEEE}, vol.~4, March
  2005, pp. 2235--2245.

\bibitem{Cooley65}
J.~M. Cooley and J.~W. Tukey, ``An algorithm for the machine calculation of
  complex fourier series,'' \emph{Math. Comp.}, vol.~19, p. 297, 1965.

\bibitem{blahut:1985}
R.~Blahut, ``Algebraic fields, signal processing, and error control,''
  \emph{Proceedings of the IEEE}, vol.~73, no.~5, pp. 874--893, May 1985.

\bibitem{ccnc}
A.~Soro and J.~Lacan, ``{FNT-based Reed-Solomon Erasure Codes},'' \emph{To
  appear in 7th Annual {IEEE} Consumer Communications and Networking
  Conference}, 2010.

\end{thebibliography}
%
%
%
%

\end{document}